\documentclass[acmsmall]{acmart}
\AtBeginDocument{%
  }

\setcopyright{acmlicensed}
\copyrightyear{2025}
\acmYear{2025}
\acmDOI{XXXXXXX.XXXXXXX}

\usepackage{geometry}
\usepackage{longtable}
\usepackage{multirow}
\usepackage{graphicx}

\begin{document}

\title{Agentic AI Software Engineers: Programming with Trust}

\author{Abhik Roychoudhury}
\authornote{Corresponding Author, Full time employment as Professor at NUS while being Senior Advisor at SonarSource.}
\email{abhik@nus.edu.sg}
\orcid{0000-0002-7127-1137}
\affiliation{%
  \institution{National University of Singapore}
  \country{Singapore}
}

\author{Corina P\u{a}s\u{a}reanu}
\email{pcorina@cmu.edu}
\orcid{0000-0002-5579-6961}
\affiliation{%
  \institution{Carnegie Mellon University, KBR Inc., NASA Ames}
  \country{USA}
}
\author{Michael Pradel}
\email{michael@binaervarianz.de}
\orcid{0000-0003-1623-498X}
\affiliation{%
  \institution{CISPA Helmholtz Center for Information Security, University of Stuttgart}
  \country{Germany}
}
\author{Baishakhi Ray}
\email{rayb@cs.columbia.edu}
\orcid{0000-0003-3406-5235}

\affiliation{%
  \institution{Columbia University}
  \country{USA}
}

\renewcommand{\shortauthors}{Roychoudhury et al.}

\begin{abstract}
Large Language Models (LLMs) have shown surprising proficiency in generating code snippets, promising to automate large parts of software engineering (SE) via artificial intelligence (AI). We argue that successfully deploying AI software engineers requires a level of trust equal to, or even greater than, the trust established by human-driven software engineering practices.
The recent trend toward LLM agents offers a path toward integrating the power of LLMs to create new code with the power of analysis tools to increase trust in the code.
This opinion piece comments on how LLM agents present AI software engineering workflows of the future, and  whether the focus of programming will shift from scale to trust.
\end{abstract}



\maketitle

\section{A Key Barrier for AI Software Engineers}

Software engineering is undergoing a disruptive phase of greater automation owing to the emergence of Large Language Models (LLMs) that generate and edit code. 
This progress creates public excitement about \emph{AI software engineers}, which promise to largely automate many core software development tasks, potentially saving tremendous costs~\cite{lancer25}.
While AI-enabled code generation and code editing are now prevalent in integrated development environments (IDEs), fully automated AI software engineers are not yet widely deployed in industrial practice. 
What is holding people back from adopting AI software engineers? 
A recent blog post by the behavioral scientist and future-of-work advocate Lindsay Kohler points out that the {\em key barrier} to AI adoption is {\em trust}~\cite{Kohler25}.
Users are wondering if they can trust AI, and how they can demonstrate trustworthiness to stakeholders.
In the domain of software engineering, the concern is thus not about the management of an organization not accepting AI software engineers, but it is about developers not trusting their new AI companions.

This brings us to the following question: 
{\em What is the place of AI software engineers in future development workflows? }
If we can figure out how automatically generated and manually written software can co-exist, this may give us a pathway of greater deployment of AI in software engineering!
Starting from early programs of just a few lines written in high-level languages in the 1960s and 70s, the size of programs has increased greatly to hundreds of millions of lines of code.
For the past fifty years, there has been a steady interest towards {\em programming in the large}. 
With the increased use of AI code generation, we believe that the emphasis will be not only on programming at scale, but increasingly on \emph{programming with trust}.

\section{Technical and Human Trust}

\begin{table}[t]
\centering
\caption{Technical and Human Angles of Trust in AI Software Engineers: What / How to Measure?}
\label{tab:trust}
\resizebox{\textwidth}{!}
{\scriptsize
\begin{tabular}{l|p{0.45\textwidth}|
p{0.4\textwidth}
}
\toprule
\textbf{Aspect} & \textbf{Trust Factor (What)} &
\textbf{Trust Factor (How)}\\
\midrule

\multirow{5}{*}{Technical} 
& Correctness: Code produce the expected results
& Reviewing \& Testing: Test the AI-generated code with sufficient and meaningful test cases \\

& Security: Code is vulnerability-free  and safe to deploy 
& Dependency analysis: Check whether the dependencies are safe, up-to-date, and  trusted \\
& Performance: Code is efficient and scalable & 
Profiling: Measure performance during execution
\\
& Maintainability: Code is easy to read, refactor, and extend &
Metrics: Measure code complexity and readability
\\
& Compliance: Code adheres to language/framework best practices &
Static analysis: Check for rule violations, e.g., with linters
\\
\midrule

\multirow{4}{*}{Human} 
& Explainability and Transparency: AI justifies the chosen solution &
Familiarity: Check whether the code is explained in ways suitable for the developer \\
& Bias and Ethics: Code reflects unbiased, ethical decisions 
& Over-reliance: Monitor and estimate if the developer is using AI coding assistance blindly \\
& Team Practice: AI aligns with the team's development workflow, experience, and review culture &
Experience mismatch: Check whether the AI generated code align with the developer’s expertise level \\

& Collaboration: Coding agent reacts well to developer feedback \\
\bottomrule
\end{tabular}}
\end{table}

Table~\ref{tab:trust} illustrates how developers perceive trust in AI-generated code, both from a technical and a human angle. 
Technical trust in AI-generated code stems from measurable attributes that determine code quality and reliability. Developers need assurance that the code is correct, secure, and performs efficiently under expected conditions. Maintainability and standards compliance are crucial for long-term usability, while thorough testing validates that the code functions as intended. Additionally, trust improves when dependencies are well-managed and sourced from reputable libraries. Human trust involves psychological and social dynamics that influence how developers perceive and accept AI-generated solutions.
A key factor is explainability and transparency: when an AI system can clarify its reasoning and design choices, developers are more likely to trust and adopt its outputs. Developers are also more likely to trust code that aligns with ethical values, follows familiar patterns and team practices, and matches the developers level of expertise.
Trust is strengthened when AI systems collaborate effectively and adapt to feedback provided by developers.

Even though we distinguish technical and human trust, we do not attempt to relate to  interpersonal trust models from sociology and psychology. 
In fact, we feel that agent-developer interactions in software will be a new mode of interaction not seen before, and new trust models may develop, e.g., building on prior efforts toward modeling trust in human-AI interactions~\cite{mehrotra2024systematic}.

\section{Software Engineering Agents}

As LLMs alone cannot inspire sufficient trust, we see \emph{LLM agents for software engineering} as a promising way of creating trustworthy AI software engineers.
What is an LLM agent for software, and how does it differ from prompt engineering?
We highlight three aspects of such agents:\footnote{See also the following for a discussion \url{ https://www.anthropic.com/research/building-effective-agents}}
\begin{itemize}
\item {\em LLMs as back-ends:} An agent is a program that leverages one or more LLMs as back-end computation and decision engines.
\item {\em Interaction with software tools:} An agent interacts with different tools to achieve a given task. In software engineering, such tools resemble those commonly used by human software engineers, e.g., file navigation, code editing, executing test suites, and invoking program analysis tools. Appropriate use of these tools is key for enhancing trust of developers in the results of the LLM agent.
\item {\em Autonomy:} An agent invokes tools in an autonomous manner. That is, the agent does not follow a deterministic algorithm, but rather creates a nondeterministic work-plan with significant autonomy.

\end{itemize}

Recently, several software engineering agents have been proposed, starting with the announcement of the Devin AI software engineer from Cognition Labs~\cite{Devin24}. Devin can solve natural language tasks (called {\em issues}), such as bug fixes and feature additions. It combines a back-end LLM with access to standard developer tools, such as a shell, a code editor, and a web browser. The agent employs such tools autonomously to let the AI software engineer mimic human practices.
In parallel with the announcement of Devin, several research groups proposed their own LLM agents for software engineering, including RepairAgent~\cite{repairagent25}, AutoCodeRover~\cite{acr24} and SWE-agent~\cite{sweagent24}.
RepairAgent~\cite{repairagent25} fixes bugs exposed by failing test cases, and guides the agent by defining a finite-state machine that outlines the typical steps followed by a developer. RepairAgent can only work with tests and cannot process natural language issues.
AutoCodeRover, a spinoff acquired by SonarSource, can solve natural language issues requiring bug fixing or feature addition. It has been integrated into the widely used SonarQube static analyzer and has already been made available to enterprise customers.
It establishes technical trust, e.g., by using program analysis on abstract syntax trees, and human trust, e.g., by extracting the intent of the software, which can be used to provide explanations of the suggested code edits. However, it makes less use of file navigation and bash tools in its implementation.
SWE-agent~\cite{sweagent24} follows a philosophy similar to Devin, by making file navigation tools and interfaces available to an AI software engineer. It does not employ any program analysis and hence cannot do intent extraction.



\section{Establishing Trust}

What makes us trust human-written code, but not necessarily the code generated by an LLM?
Part of the reason is the perceived capability of ``passing the blame''.
If a human developer is involved, there is the promise of getting feedback from the developer as needed.
Of course, this does not always hold, e.g., if the developer eventually leaves an organization. Nevertheless, accepting a code commit from a developer partially depends on the reputation of the developer within the organization.
For an AI software engineer in the form of an LLM agent to earn a reputation similar to a senior human colleague, it has to integrate established quality assurance techniques and collaborate effectively with human developers. We outline several ideas toward this goal.

\paragraph*{Testing and Lightweight Static Analysis}
One way to increase technical trust is to retrofit testing and lightweight static analyzers into an LLM agent.
For example, in the process of code generation, additional artifacts, such as tests that exercise the newly added code, can be generated as well~\cite{ryan2024code}, possibly derived from a natural language description of the agent's task.
A crucial challenge is to create appropriate test oracles that check the actual outputs against expected outputs, e.g., by inferring the expected outputs from natural language specifications.

\paragraph*{Formal Proofs}
An enhanced degree of technical trust can come from formal proofs. A promising paradigm  in this regard is automated, proof-oriented programming~\cite{POP}.
In this paradigm, LLMs generate the code together with the necessary formal specifications (pre/post-conditions, loop invariants, and so on) in a verifiable language, such as F*, Dafny, or Verus.
Such programs can then be automatically verified, providing greater trust than testing alone.

\paragraph*{Guardrails for Increased Security and Alignment}
Trust can also be ensured through the use of {\em guardrails}. 
These can serve as a sanitization mechanisms, filtering malicious inputs before they reach the LLM and validating the generated code before it is returned to a user. 
Guardrails help defend against three key threats: prompt injections, where prompts trick the LLM into bypassing safety measures; malicious code, where harmful input code leads the LLM to generate or modify malicious code; and vulnerable code, where even unintentionally insecure input may cause the LLM to propagate or alter undesired code.

\paragraph*{Specification Inference for Explainability}
A more conceptual mechanism to establish both technical and human trust would be to infer the code intent from the initial, possibly buggy program.
The system-level intent of what a large software system is supposed to do can often be crisply captured by a detailed natural language prompt.
What is missing is the intent of the  functions or methods.
An LLM agent could be geared towards such specification inference, navigating the code base via code search, and trying to infer the intended behavior~\cite{acr25}.
Such explicit unit-level  specification inference can enable the program modifications  to be accompanied by justifications.



\paragraph*{Effective AI-Human Collaboration}
A key factor for increasing human trust into AI software engineers will be to enforce effective AI-human collaboration patterns.
As experienced first-hand by the AutoCodeRover team~\cite{acr24} via real-life anecdotes communicated by clients, developer hesitation in accepting AI-generated code also comes from the volume of code that can be quickly generated by AI tools, overwhelming human developers. Providing AI generated code with {\em confidence scores} can  reduce developer hesitation. Validating (or refuting) such anecdotal experiences via studies on how to reduce developer hesitation remains a direction of future research. Latest release of OpenAI's Codex agent 
in May 2025 (the Codex agent can run synchronously in the terminal watching a human, while also running long-running tasks in the cloud asynchronously) sharpens these questions -- can a LLM coding agent be trusted? To operationalize socio-technical integration for AI agent trust, organizations should implement review parity—requiring AI-generated code changes to undergo identical peer review processes as human contributions, e.g., by enforcing the same two-reviewer gates and quality thresholds. Additionally, deploying specialized code review agents backed by trusted static and dynamic analysis tools may assist human reviewers. In the development workflow, organizations should incorporate transparent scaffolding by systematically tagging AI-generated code, embedding authorship provenance and confidence metrics directly within pull request templates and review interfaces. In this way, trust measures can be integrated inside software workflows.

\section{Outlook}

As AI software engineers take on more core development tasks, their success will hinge not just on technical capability but on earning developer trust.
Rather than relying on many separate agents for specialized software engineering tasks, it could be worthwhile to create a unified software engineering agent that combines coding, testing, debugging, etc.\ into a coherent, explainable workflow.
To become a trusted collaborator, such an agent must offer transparency, adapt to feedback, and integrate safeguards that ensure quality and security. Programming with AI will mean not full automation, but effective delegation, where human and AI work hand in hand.


\bibliographystyle{ACM-Reference-Format}
\bibliography{main}


\end{document}